\documentclass[preprint,a4paper,5p,times,twocolumn,sort&compress]{elsarticle}

\usepackage{amssymb}

\usepackage[utf8]{inputenc}
\usepackage{graphicx}
\usepackage{hyperref}
\usepackage{float}
\usepackage{amsmath}
\usepackage{amssymb}
\usepackage[T1]{fontenc}
\usepackage{hyperref} 
\usepackage{isotope} 
\usepackage{upgreek}
\usepackage{sidecap}
\usepackage{wrapfig}
\usepackage{placeins}
\usepackage{subcaption}
\usepackage{booktabs}
\usepackage{color}

\usepackage{lineno}

\def\cd113{\isotope[113][]{Cd}}
\def\gA{$g_{\rm A}$}

\journal{Physics Letters B}

\begin{document}

\begin{frontmatter}



\title{Quenching of $g_{\rm A}$ deduced from the $\beta$-spectrum shape of $^{113}$Cd\\ measured with the COBRA experiment}


\author[dortmund]{Lucas~Bodenstein-Dresler}
\author[dresden]{Yingjie~Chu}
\author[dresden]{Daniel~Gehre}
\author[dortmund]{Claus~G\"o{\ss}ling}
\author[dresden]{Arne~Heimbold}
\author[dortmund]{Christian~Herrmann}
\author[prague]{Rastislav~Hodak}
\author[jyvaskyla]{Joel~Kostensalo}
\author[dortmund]{Kevin~Kr\"oninger}
\author[dresden]{Julia~K\"uttler}
\author[dortmund]{Christian~Nitsch}
\author[dortmund]{Thomas~Quante}
\author[prague]{Ekaterina~Rukhadze}
\author[prague]{Ivan~Stekl}
\author[jyvaskyla]{Jouni~Suhonen}
\author[dortmund]{Jan~Tebr{\"u}gge}
\author[dortmund]{Robert~Temminghoff}
\author[dresden]{Juliane~Volkmer}
\author[dresden]{Stefan~Zatschler\corref{corauthor}}
\cortext[corauthor]{corresponding author}
\ead{stefan.zatschler@tu-dresden.de}

\author[dresden]{Kai~Zuber}

\address{\rm\normalsize COBRA collaboration}

\address[dortmund]{TU Dortmund, Lehrstuhl f\"ur Experimentelle Physik IV, Otto-Hahn-Str.~4a, 44221 Dortmund, Germany}
\address[dresden]{TU Dresden, Institut f\"ur Kern- und Teilchenphysik, Zellescher Weg 19, 01069 Dresden, Germany}
\address[prague]{CTU Prague, Institute of Experimental and Applied Physics, CZ-11000 Prague, Czech Republic}
\address[jyvaskyla]{University of Jyvaskyla, Department of Physics, P.O. Box 35, FI-40014, Finland}


\begin{abstract}
A dedicated study of the quenching of the weak axial-vector coupling strength $g_{\rm A}$ in nuclear processes has been performed by the COBRA collaboration.
This investigation is driven by nuclear model calculations which show that the $\beta$-spectrum shape of the fourfold forbidden non-unique decay of \cd113 strongly depends on the effective value of $g_{\rm A}$.
Using an array of CdZnTe semiconductor detectors, 45 independent \cd113 spectra were obtained and interpreted in the context of three nuclear models.
The resulting effective mean values are $\overline{g}_{\rm A}(\text{ISM}) = 0.915 \pm 0.007$, $\overline{g}_{\rm A}(\text{MQPM}) = 0.911 \pm 0.013$ and $\overline{g}_{\rm A}(\text{IBFM-2}) = 0.955 \pm 0.022$.
These values agree well within the determined uncertainties and deviate significantly from the free value of $g_{\rm A}$.
This can be seen as a first step towards answering the long-standing question regarding quenching effects related to $g_{\rm A}$ in low-energy nuclear processes.
\end{abstract}


\end{frontmatter}


\section{Introduction}
The potential quenching of the weak axial-vector coupling strength $g_{\rm A}$ in nuclei is of general interest, e.g. in nuclear astrophysics, rare single $\beta$-decays as well as double $\beta$-decays.
The predicted rate for neutrinoless double beta ($0\nu\beta\beta$) decay, in particular in the case of light Majorana neutrino exchange, depends strongly on the numerical value of $g_{\rm A}$ through the leading Gamow-Teller part of the $0\nu\beta\beta$ nuclear matrix element (NME).
A wide set of nuclear-theory frameworks has been adopted to calculate the value of this NME \cite{REPORT,Suhonen2012c,Suhonen2012d,Engel2017,Ejiri2019} but the associated quenching of $g_{\rm A}$ has not been addressed quantitatively.
The effective value of $g_{\rm A}$ can be considerably quenched at least in low-energy processes such as single $\beta$-decays and two-neutrino double beta ($2\nu\beta\beta$) decays \cite{Juoda2005,Faess2008,Cau2012,Suh2013c,Suh2014,Eji2014,Eji2015,Pir2015,Dep2016}.
This quenching can strongly affect the sensitivity of the presently running $0\nu\beta\beta$-experiments \cite{Suhonen2017a}
including GERDA \cite{Agostini2018} and the 
\mbox{MAJORANA} DEMONSTRATOR \cite{Aalseth2018} ($^{76}$Ge),  
NEMO-3 \cite{Arnold2004,Argyriades2010,Arnold2015,Arnold2017} ($^{82}$Se, $^{96}$Zr, $^{100}$Mo, $^{116}$Cd), 
COBRA \cite{Zatschler2015} ($^{116}$Cd),
CUORE \cite{Alduino2018} ($^{130}$Te),
EXO-200 \cite{Albert2018} and KamLAND-Zen \cite{Gando2016} ($^{136}$Xe)
and future projects such as LEGEND \cite{Legend2017} ($^{76}$Ge),
SuperNEMO \cite{Hodak2015}, 
AMoRE \cite{Park2015} and LUMINEU \cite{Becker2016} ($^{100}$Mo), 
MOON \cite{Fushimi2010} ($^{82}$Se, $^{100}$Mo),
AURORA \cite{Barabash2018} ($^{116}$Cd), 
SNO+ \cite{Andringa2016} and  CUPID \cite{Artusa2017} ($^{130}$Te), 
NEXT-100 \cite{Martin-Albo2016} as well as nEXO \cite{Albert_nEXO_2018} and 
\mbox{PandaX-III} \cite{Chen2017}~($^{136}$Xe).
Since $0\nu\beta\beta$-decay is a high-momentum exchange process of $\sim$100\,MeV it is not clear how the results obtained for the quenching of $g_{\rm A}$ in the low-momentum exchange single \mbox{$\beta$-decays} and $2\nu\beta\beta$-decays can be translated to $0\nu\beta\beta$-decay.
Nevertheless, the conversion from the potentially measured $0\nu\beta\beta$-decay half-lives into a Majorana neutrino mass has a strong $g_{\rm A}$ dependence due to the involved Gamow-Teller NME. This is why it is important to study the quenching of $g_{\rm A}$ in as many ways as possible, even in low-energy processes, like single \mbox{$\beta$-decays}.
The quenching of $g_{\rm A}$ at low energies has several different sources: Non-nucleonic degrees of freedom (e.g. delta resonances) and giant multipole resonances (like the Gamow-Teller giant resonance) removing transition strength from low excitation 
energies.
Further sources of quenching (or sometimes enhancement, see \cite{Kostensalo2018}) are nuclear processes beyond the impulse approximation (in-medium meson-exchange or two-body weak currents) and deficiencies in the handling of the nuclear many-body problem (too small single-particle valence spaces, lacking many-body configurations, omission of three-body nucleon-nucleon interactions, etc.).
Different methods have been introduced to quantify the quenching effect in decay processes of low momentum exchange (see the review \cite{Suhonen2017b}).
One method recently proposed exploits the dependence of the $\beta$-spectrum shape of highly-forbidden non-unique decays on $g_{\rm A}$.
This approach will be introduced in the following.

\subsection{The spectrum-shape method}
\label{sec:SSM}
In Ref.~\cite{Haaranen2016} it was proposed that the shapes of \mbox{$\beta$-electron} spectra could be used to determine the values of the weak coupling strengths by comparing the shape of the computed spectrum with the measured one for forbidden non-unique $\beta$-decays.
This method was coined the spectrum-shape method (SSM) and its potential in
determining the values of the weak coupling strengths $g_{\rm V}$ (vector part) and $g_{\rm A}$ (axial-vector part) is based on the complexity of the \mbox{$\beta$-electron} spectra.
The corresponding $\beta$-decay shape factor $C(w_{e})$, $w_e$ being the total energy of the emitted electron ($\beta^-$-decay) or positron ($\beta^+$-decay) in units of $m_e$, is an involved combination of different NMEs and phase-space factors \cite{Mustonen2006} and can be decomposed \cite{Haaranen2016} into vector, axial-vector and mixed vector-axial-vector parts in the form 
\begin{equation}
C(w_e)= g_{\rm A}^2\Big\lbrack C_{\rm A}(w_e) + \frac{g_{\rm V}}{g_{\rm A}}C_{\rm VA}(w_e) + 
\left(\frac{g_{\rm V}}{g_{\rm A}}\right)^2C_{\rm V}(w_e)\Big\rbrack .
\label{eq:decomp}
\end{equation}
In Ref.~\cite{Haaranen2016} also the next-to-leading-order corrections to $C(w_{e})$ were included.
In the same reference it was noticed that the \mbox{$\beta$-spectrum} shape for the fourfold forbidden non-unique ($\Delta J^\pi$ = $4^+$) ground-state-to-ground-state $ \beta^-$-decay branch \mbox{$^{113}\mathrm{Cd}(1/2^+)\to\,^{113}\mathrm{In}(9/2^+)$} is highly sensitive to the ratio $g_{\rm V}/g_{\rm A}$ in Eqn.~(\ref{eq:decomp}), and hence a comparison of the calculated spectra with the one measured by e.g. Belli \textit{et~al.} \cite{Belli2007} could open a way to determine the value of this ratio.
In Ref.~\cite{Haaranen2016} the theoretical electron spectra were computed by using the microscopic quasiparticle-phonon model (MQPM) \cite{Toivanen1998} and the interacting shell model (ISM).
This work was extended in \cite{Haaranen2017} to include a comparison with the results of a third nuclear model, the microscopic interacting boson-fermion model (IBFM-2) \cite{Iachello1991}.
The studies \cite{Haaranen2016,Haaranen2017} were continued by the works \cite{Kostensalo2017b} and \cite{Suhonen2017c} where the evolution of the $\beta$-spectra with changing value of $g_{\rm V}/g_{\rm A}$ was followed for a number of highly-forbidden \mbox{$\beta^-$-decays} of odd-$A$ nuclei (MQPM and ISM calculations) and even-$A$ nuclei (ISM calculations). 

There are also some potential uncertainties related to the SSM. One problem is the delicate balance of the vector, axial-vector and mixed vector-axial-vector parts in Eqn.~(\ref{eq:decomp}) in the range where the SSM is most sensitive to the ratio $g_{\rm V}/g_{\rm A}$. 
At this point one has to rely on results which require cancellations at a sub-percent level (see the review \cite{Suhonen2017b}).
On the other hand, this point of cancellation seems to be similar for different nuclear
models and quite insensitive to the parameters of the adopted model Hamiltonians and the details of the underlying mean field.
Nevertheless, quantification of the associated uncertainties is non-trivial and here we estimate the systematic uncertainty in the extracted values of $g_{\rm A}$ (assuming vector-current conservation, $g_{\rm V}=1$) by using three different nuclear-model frameworks (ISM, MQPM, IBFM-2) in our computations.
One particular problem of the present calculations is that the used nuclear models cannot predict the half-life of \mbox{$^{113}\mathrm{Cd}$} and the electron spectral shape for consistent values of $g_{\rm A}$ and $g_{\rm V}$.
This was already pointed out in Ref.~\cite{Haaranen2016} and further elaborated in Ref. \cite{Haaranen2017}.
The reason for this could be associated with the deficiencies of the adopted nuclear Hamiltonians in the presently discussed nuclear mass region $A\sim 110$ and/or a need for a more nuanced treatment of the effective renormalization of the weak coupling constants, separately for different transition multipoles, like done in the context of first-forbidden non-unique transitions (see the examples in the review \cite{Suhonen2017b}).
One has also to bear in mind that the half-life depends on the values of both $g_{\rm A}$ and $g_{\rm V}$ whereas the normalized spectrum shape depends only on the ratio of them. Thus the SSM can be used to fix the ratio $g_{\rm V}/g_{\rm A}$ whereas the half-life can be used to fix the absolute value of e.g. $g_{\rm A}$.

\subsection{Previous studies on \isotope[113]{Cd}}
\label{sec:previous_Cd113}

\begin{table*}[ht!]
\centering
\caption{Summary of the most important previous \cd113 studies. Listed are the detection threshold $E_\text{th}$, the isotopic exposure for \cd113, the energy resolution quoted as FWHM at the accepted AME2016 $Q$-value \cite{AME2016}, the signal-to-background ratio as well as the experimentally determined half-life $T_{1/2}$. Statistical and systematic uncertainties were added in quadrature, if quoted separately.}
\label{tab:previous_studies}
\begin{tabular}{ccccccc}
\toprule
Detector material & $E_\text{th}$ / keV & isotop. exp. / kg\,d & FWHM / keV & S/B ratio & $T_{1/2}$ / $10^{15}$\,yrs & Ref., year \\
\midrule
CdWO$_4$, 454\,g			& 44		& 0.31	& $\sim$ 49 & $\sim$50 	& 7.7$\pm$0.3 & \cite{Danevich1996}, 1996\\
CdZnTe, 3$\times$5.9\,g		& 100	& 0.05	& $\sim$ 43 & $\sim$ 8	& 8.2$_{-1.0}^{+0.3}$ & \cite{Cobra_Cd113_2005}, 2005\\
CdWO$_4$, 434\,g			& 28 		& 1.90 	& $\sim$ 47	& $\sim$56  & 8.04$\pm$0.05 & \cite{Belli2007}, 2007\\
CdZnTe, 11$\times$6.5\,g	& 110 		& 0.38 	& $\sim$ 20	& $\sim$ 9	& 8.00$\pm$0.26 & \cite{Cobra_Cd113_2009}, 2009\\
CdZnTe, 45$\times$6.0\,g	& 84 		& 2.89	& $\sim$ 18	& $\sim$47	& - & present work\\
\bottomrule
\end{tabular}
\end{table*}

The fourfold forbidden non-unique $\beta$-decay of \isotope[113]{Cd} was studied before using different experimental techniques.
The main focus was to determine its \mbox{$Q$-value} and half-life.
Among them are low-background experiments using CdWO$_4$ scintillator crystals and CdZnTe semiconductor detectors like the \mbox{COBRA} experiment \cite{Zuber2001}.
A summary of the most recent studies is given in Tab.~\ref{tab:previous_studies}.
The most precise half-life measurement was achieved with a CdWO$_4$ scintillator in 2007 \cite{Belli2007}.
The same CdWO$_4$ crystal was already used ten years before in a similar study \cite{Danevich1996}.
CdWO$_4$ scintillators reach typically lower thresholds, but feature a worse energy resolution compared to CdZnTe solid state detectors as used for COBRA.
Additionally, the \cd113 $\beta$-decay was investigated with early predecessors of the current COBRA demonstrator \cite{Cobra_Cd113_2005, Cobra_Cd113_2009}.
The latter study resulted in a half-life of \mbox{$(8.00 \pm 0.11\ (\text{stat.}) \pm 0.24\ (\text{syst.}))\times 10^{15}$ years} and a $Q$-value of $322.2 \pm 0.3\ (\text{stat.}) \pm 0.9\ (\text{syst.})\,$keV.
It is noteworthy that this $Q$-value is in perfect agreement with the accepted AME2016 value of $Q_\beta = 323.83 \pm 0.27$\,keV \cite{AME2016} while it is several 10\,keV off for Ref.~\cite{Belli2007}.
The other studies listed in Tab.~\ref{tab:previous_studies} do not include an experimentally determined $Q$-value.

In addition, first attempts to describe the $\beta$-spectrum shape with conventional shape factors were pursued.
All previous studies assumed that the \cd113 \mbox{$\beta$-decay} can be described approximately with a shape factor corresponding to a threefold forbidden unique decay \mbox{($\Delta J^\pi$ = $4^-$)}. This is a clear oversimplification probably due to the lack of accurate calculations at that time.
Furthermore, the extracted shape factors are inconclusive as pointed out in Ref. \cite{Belli2007} and \cite{Cobra_Cd113_2009}.
The authors of Ref.~\cite{Belli2007} already mentioned that there is a discrepancy between the assumed polynomial fit and the experimental spectrum above 250\,keV, if the $Q$-value is fixed to the accepted value quoted above.
Nowadays, there is no justification to assume such an oversimplified parametrization.
Instead, the present work is based on the SSM using calculations of the full expression for transitions with $\Delta J^\pi = 4^+$.

More recently, the COBRA collaboration used the \cd113 \mbox{$\beta$-decay} to investigate the demonstrator's detector stability by monitoring the average decay rate over the time scale of several years~\cite{Cobra_stability_2016}.
Such a continuous data-taking was not achieved in any experiment dedicated to the study of long-lived $\beta$-decays before.
The analysis threshold for this particular study was $\sim$170\,keV, which is comparatively high.
It was chosen to ensure that the stability study is representative for the whole energy range interesting for $\beta\beta$-decay searches with COBRA excluding noise near the detection threshold.
On the other hand, this drastically limits the available \cd113 energy range.
Following this study, modifications on the hardware and software level were made to optimize the demonstrator setup for a dedicated low-threshold run with the aim to investigate the \cd113 $\beta$-electron spectrum shape with high precision.

In this article we present the results of a dedicated \cd113 measurement campaign with the COBRA demonstrator.
This study features the best signal-to-background ratio of all previous CdZnTe analyses, high statistics, a good energy resolution and moderate thresholds while providing 45 independent \mbox{$\beta$-spectra} of the transition \mbox{$^{113}\mathrm{Cd}(1/2^+)\to\,^{113}\mathrm{In}(9/2^+)$}.
The data will be used to evaluate quenching effects of $g_{\rm A}$ in the context of the three nuclear models (ISM, MQPM, IBFM-2) using the SSM as introduced in section \ref{sec:SSM}.

\section{Experimental setup}
The COBRA collaboration searches for $\beta\beta$-decays with room temperature CdZnTe (CZT) semiconductor detectors.
As \mbox{$0\nu\beta\beta$-decay} is expected to be an extremely rare process, the experiment is located at the Italian Laboratori Nazionali del Gran Sasso (LNGS), which is shielded against cosmic rays by 1400\,m of rock.
Currently, it comprises 64 coplanar-grid (CPG) detectors arranged in four layers of $4\times4$ crystals.
This stage of the experiment is referred to as the \mbox{COBRA} demonstrator \cite{Cobra_demonstrator_2016}.
Each crystal has a size of about 1$\times$1$\times$1\,cm$^3$ and a mass of about $6.0\,$g.
All of them are coated with a clear encapsulation known to be radio-pure.
In previous iterations of the experiment it was found that the formerly used encapsulation lacquer contained intrinsic contaminations on the order of 1\,Bq/kg for the long-lived radio-nuclides \isotope[238]{U}, \isotope[232]{Th} and \isotope[40]{K}.
The new encapsulation lacquer has been investigated with \mbox{ICP-MS} at the LNGS, which confirmed the improved radio-purity with determined specific activities on the order of 1\,mBq/kg for \isotope[238]{U} and \isotope[232]{Th} and about 10\,mBq/kg for \isotope[40]{K}.
The four layers are framed by polyoxymethylene holders which are installed in a support structure made of electroformed copper.
The inner housing is surrounded by 5\,cm of electroformed copper, followed by 5\,cm of ultra-low activity lead ($<3$\,Bq/kg of \isotope[210][]{Pb}) and 15\,cm of standard lead.
Additionally, the inner part is enclosed in an air-tight sealed box of polycarbonate, which is constantly flushed with evaporated nitrogen to suppress radon-induced background.
Outside the inner housing the first stage of the read-out electronics is located.
The complete setup is enclosed by a construction of iron sheets with a thickness of 2\,mm, which acts as a shield against electromagnetic interferences.
The last part of the shielding is a layer of 7\,cm borated polyethylene with 2.7 wt.\% of boron to effectively suppress the external neutron flux.

Charge-sensitive pre-amplifiers integrate the current pulses induced by particle interactions with the sensitive detector volume and convert the single-ended detector pulses into differential signals in order to minimize electronic noise during transmission.
After linear amplification, the pulse shapes are digitized using $100\,$MHz flash analog-to-digital converters (FADCs) with a sample length of 10\,$\mu$s.
Each FADC has eight input channels allowing for the read-out of four CPG detectors with two anode signals each.
The clock speed and potential offset of the individual FADCs are corrected with the help of artificial pulses injected by a generator into the data acquisition chain.
These are processed like real detector signals and provide well-defined synchronization points for an offline synchronization of the data.
After this it is possible to identify and reject multi-detector hits for single detector analyses.
The achievable accuracy of the time synchronization is about 0.1\,ms. 
Additional key instruments in background suppression for COBRA are the reconstruction of the so-called interaction depth \cite{Fritts2013} and the use of pulse-shape discrimination techniques \cite{Fritts2014, Zatschler2016}.
The interaction depth $z$ is referred to as the normalized distance between the anode grid ($z=0$) and the planar cathode ($z=1$).

\section{Data-taking and event selection}
\subsection{Run preparation}
\label{sec:LNGS_preparation}
In preparation of a dedicated \cd113 run, the potential of optimizing the COBRA demonstrator towards minimum threshold operation was studied in detail.
One major improvement was achieved by exchanging the coolant in the cooling system of the pre-amplifier stage, which allows operation at lower temperatures.
The direct cooling of the first stage of the electronics dramatically reduces the thermal component of the signal noise while at the same time the detector performance benefits from an ambient temperature slightly below room temperature.
The temperature inside the inner shield of the experiment is monitored by several sensors at different positions.
In agreement with previous studies on CPG-CZT detectors \cite{Dawson_Cobra_2009} an optimal temperature was found to be around 9$^\circ$C.
The crystals themselves are not cooled directly, but through convection and radiation cooling they are kept at the same temperature as the surrounding shielding components.
For each temperature set the optimal trigger threshold for every channel had to be determined after reaching the thermal equilibrium.
This was done by monitoring the average trigger rate on a daily basis and adjusting the individual thresholds accordingly.
While accomplishing this optimization, the worst-performing detector channels were switched off to prevent potential sources of electromagnetic interferences and crosstalk.
The COBRA demonstrator was then calibrated at the point of best performance and the dedicated \cd113 data-taking period was started.
It lasted from Jul.'17 until Feb.'18. 
During this period, the individual trigger thresholds $E_\text{th}$ were mostly kept at the same level resulting in an average of $\overline{E}_\text{th} = 83.9\pm 14.8$\,keV considering the average energy resolution in terms of FWHM at this energy.
It should be noted that this is not the minimum amount of energy $E_0$ that can be measured by the detectors, but includes a correction function $f_\text{cor}(z)$ depending on the interaction depth $z$ to ensure that the spectrum shape is not distorted by the event reconstruction (see Ref. \cite{Fritts2013}).
\begin{eqnarray}
E_\text{th} = E_0 \cdot f_\text{cor}(z), \quad \text{with } f_\text{cor}(0) \approx 1.6.
\label{eqn:thresh}
\end{eqnarray}
Furthermore, an analysis threshold $\tilde{E}_\text{th}$ is introduced by modifying Eqn.~(\ref{eqn:thresh}) to $\tilde{E}_\text{th} = E_\text{th}+8\,$keV as will be motivated in section~\ref{sec:shape_comparison}.
Such a careful and conservative threshold correction is not discussed in the previous studies summarized in Tab.~\ref{tab:previous_studies}.
The individual detector thresholds $E_\text{th}$ range from 44\,keV to 124\,keV, whereas the 18 best detectors were operated below or around 70\,keV and only the four worst-performing at 124\,keV.
For comparison, the threshold quoted in Ref.~\cite{Belli2007} using a CdWO$_4$ scintillator can be referred to as $28\pm 14$\,keV considering the given energy resolution.
This is not far away from what has been achieved in the present study, where in addition a much higher number of detector channels could be used.

\subsection{Detector calibration and characterization}
\label{sec:calibration}
The energy calibration of each detector was done using the radio-nuclides \isotope[22][]{Na}, \isotope[152][]{Eu} and \isotope[228][]{Th} providing \mbox{$\gamma$-lines} in the range from 121.8\,keV to 2614.5\,keV.
Each line was fitted with a Gaussian plus a polynomial function to take into account the underlying Compton continuum.
The calibration is done by a linear fit of the peak position in channel numbers versus the known $\gamma$-line energy.
Using the fit results, the energy resolution quoted as full-width at half-maximum (FWHM) can be parametrized for each detector separately as
\begin{eqnarray}
\text{FWHM}(E) = \sqrt{p_0 + p_1\cdot E + p_2\cdot E^2}, \ p_i > 0.
\label{eqn:FWHM}
\end{eqnarray}
The parameter $p_0$ is independent of the deposited energy and accounts for a constant contribution from noise.
The second term scales with $\sqrt{E}$ and is motivated by the Poisson fluctuations of the charge carrier production while the third term is a rather small correction for detector effects.
With the parametrization given in Eqn.~(\ref{eqn:FWHM}) the achieved mean relative resolution FWHM$(E)/E$ ranges from $12.5\pm0.6$\% at 121.8\,keV to $1.7\pm0.1$\% at 2614.5\,keV including the uncertainty on the mean.
The spread of the FWHM$(E)/E$ distribution can be expressed by the according standard deviations of 4.0\% (121.8\,keV) and 0.7\% (2614.5\,keV).

\subsection{Detector pool selection}
To ensure stable operation during the dedicated \cd113 run, data with only a subset of the 64 installed detectors was collected.
Three detectors were known to suffer from problems with the data acquisition electronics and unreliable contacting.
Those were switched off from the very beginning.
In addition, twelve detectors, that had to be operated with a threshold higher than 200\,keV, were switched off subsequently during the setup optimization.
Four additional detectors showed an altered performance comparing the results of calibration measurements during the \cd113 run.
The data of those were excluded from the final spectrum shape analysis.
In the end, 45 out of 49 operated detectors qualified for the analysis with an average exposure of 1.10\,kg\,d per detector.
Two detectors were partly disabled and feature a reduced exposure of 0.29\,kg\,d and 0.79\,kg\,d, respectively.
Using the natural abundance of \cd113 of \mbox{12.225\% \cite{IUPAC2013}} in combination with the molar mass fraction of cadmium in the detector material, referred to as Cd$_{0.9}$Zn$_{0.1}$Te, it follows that the \cd113 isotopic exposure makes up for 5.84\% of the total exposure.
The combined isotopic exposure of all selected detectors adds up to 2.89\,kg\,d.

\subsection{Event selection}
The standard COBRA selection cuts are used in the \cd113 analysis (see \cite{Cobra_limits_2016} for reference).
Firstly, coincidences between all operational detectors are rejected, which is possible after synchronizing the 16 FADC clocks.
The coincidence time window used to declare two events as simultaneous is set to 0.1\,ms.
The experiment's timing accuracy is about a factor of 30 better than achieved in previous studies (e.g. Ref.~ \cite{Belli2007}, 3.16\,ms), which minimizes the loss of events due to random coincidences caused by potential \cd113 decays in different source crystals.
The next stage consists of a set of data-cleaning cuts (DCCs) to remove distorted and unphysical events.
The validity of those cuts was checked with a special run, where all channels of the same FADC were read out, if the trigger condition was fulfilled for a single channel.
The triggered event trace was then rejected and only the remaining baseline pulses were analyzed.
Those pulses were treated as a proxy for noise-only signals.
It was found that $99.8\pm0.1$\% of the untriggered events are rejected by the DCCs while there is no significant variation between single channels.
The signal acceptance of the DCCs is sufficiently constant over the \cd113 energy range and has been determined to $87.5\pm0.6$\%.
After applying the DCCs the remaining events of the noise-only data are limited to energies $< 40$\,keV, which is well below the anticipated analysis threshold.
Part of the DCCs is also a mild cut on the interaction depth $z$ to remove near-anode reconstruction artifacts.
The interaction depth is further restricted to remove events with an unphysically high $z$.
The depth selection is optimized for each detector individually and covers, for the majority, the range \mbox{$0.2 < z \leq 0.97$}.
No further pulse-shape discrimination cuts are necessary since the \cd113 decay is by far the strongest signal for COBRA at low energies.

\subsection{Background description}
\label{sec:background}
\begin{figure}[ht!]
\centering
\includegraphics[width=0.5\textwidth]{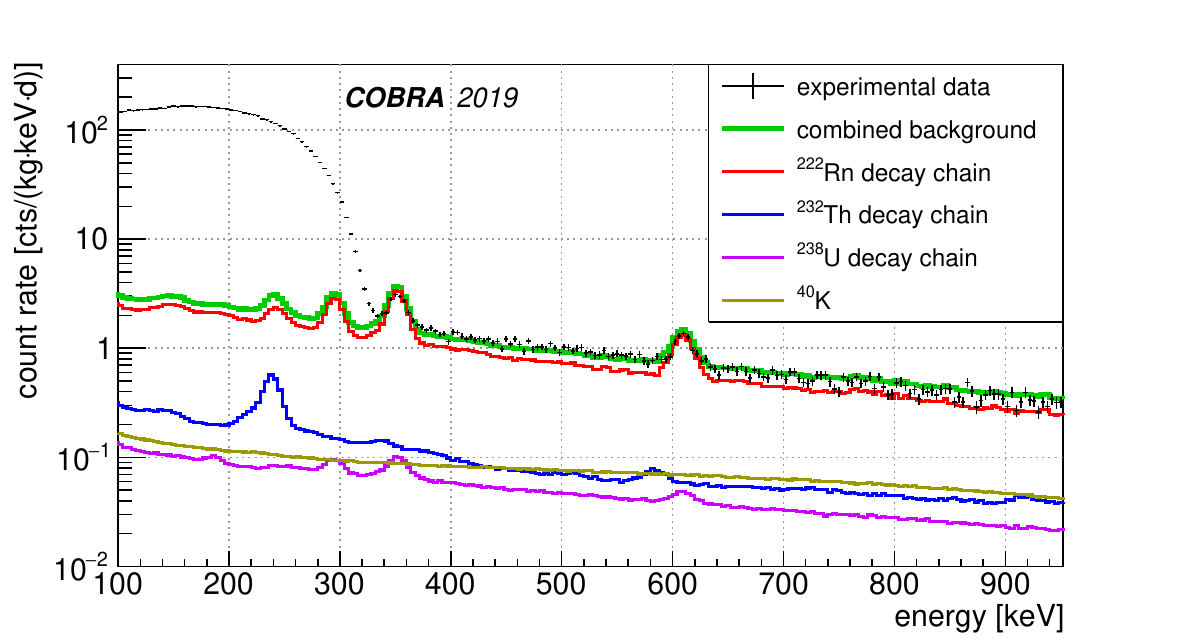}
\caption{Combined data of all detectors compared to the MC background prediction for the \cd113 energy range. Considered are the \isotope[222]{Rn}, \isotope[232]{Th} and \isotope[238]{U} decay chains as well as \isotope[40]{K} with different origins of the primary decays.}
\label{fig:background_model}
\end{figure}

Above the \cd113 $Q$-value, the measured count rate drops by at least two orders of magnitude.
Compared to the previous COBRA study \cite{Cobra_Cd113_2009}, this is an improvement of about one order magnitude.
The maximum count rate for the combined \cd113 spectrum of all detectors (see Fig.~\ref{fig:background_model}) is about 175\,cts/(kg\,keV\,d) at 150\,keV and drops sharply to below 1.5\,cts/(kg\,keV\,d) at 400\,keV.
The background decreases exponentially for higher energies and is studied up to $\sim$10\,MeV.

Previous CdWO$_4$ studies limited the background study to much lower energies, e.g. in Ref. \cite{Belli2007} for a first background run up to 1.7\,MeV and for the \cd113 data-taking to 0.6\,MeV.
In the high-energy region two $\alpha$-decay peaks are present originating from \isotope[190]Pt ($Q_\alpha  = 3.2$\,MeV) and \isotope[210]Po ($Q_\alpha  = 5.4$\,MeV).
Platinum is part of the electrode metalization while \isotope[210]Po is a daughter nucleus of the radon decay chain.
Both event populations are removed completely by a cut on the interaction depth due to their localized origin on the cathode side (see e.g. Ref.~\cite{Zatschler2015}, Fig.~2). 
Nonetheless, surface contaminations with radio-nuclides, especially from the radon decay chain, are found to be the dominating source of background for the search for \mbox{$0\nu\beta\beta$-decay}.
In the most recent $0\nu\beta\beta$-decay search of \mbox{COBRA}, the background index for the \isotope[116]{Cd} ($Q_{\beta\beta}=2.8\,$MeV) region of interest (ROI) is quoted as 2.7\,cts/(kg\,keV\,yr) \cite{Cobra_limits_2016}.
The background in this energy range is expected to be dominated by $\alpha$-decays on the lateral surfaces.
The $\alpha$-particles have to pass through the encapsulation lacquer of about 20\,$\mu$m thickness before they can deposit energy in the sensitive detector volume.
Because of the inhomogeneity of the lacquer, the according $\alpha$-spectrum is strongly deteriorated without a noticeable peak position.
Near the \cd113 ROI there are only two prominent \mbox{$\gamma$-lines} visible in the combined spectrum of all detectors.
These lines originate from the decays of \isotope[214]Pb (351.9\,keV) and \isotope[214]Bi (609.3\,keV) as short-living \isotope[222]Rn daughters and correspond to the dominant de-excitation processes.
In Dec.'17 there was a short period without nitrogen flushing of the inner shield of the experiment contributing to the overall radon exposure.
Nevertheless, the effect on the spectrum shape is completely negligible since the \mbox{$\gamma$-lines} only produce weak Compton continua.
The background contribution to the \cd113 region is estimated with a Monte Carlo (MC) simulation based on GEANT4 \cite{Allison2016} using the \textit{shielding} physics list, which is recommended for low-background experiments.
The background projection considers the \isotope[222]{Rn} decay chain within the gas layer of the geometry and near-detector contaminants from the \isotope[238]{U} and \isotope[232]{Th} decay chains as well as \isotope[40]{K}.
Impurities of the primordial radio-nuclides can only contribute marginally to the background due to the observed absence of characteristic prompt $\gamma$-lines, such as the de-excitation lines of \isotope[208]{Tl} at 583.2\,keV and 2614.5\,keV.
The signal-to-background ratio can be calculated as the integral over the \cd113 ROI, defined by the average threshold of $\overline{E}_\text{th} = 83.9$\,keV and the $Q$-value, for the experimental data and MC prediction.
This leads to $S/B \approx 46.8$, which is comparable to former CdWO$_4$ studies and about a factor of five better than previous COBRA studies (see Tab.~\ref{tab:previous_studies}). 
It should be noted that the background composition for the dedicated \cd113 run is different compared to the latest $0\nu\beta\beta$-decay analysis using data of the same setup from Oct.'11 to Sept.'16 \cite{Cobra_limits_2016}.
There is no indication for the previously observed annihilation line at 511\,keV or the \isotope[40]{K} $\gamma$-line at 1460.8\,keV.
One reason for this is that no pulse-shape discrimination cuts are used in the present analysis because the efficiency of those is rather poor at low energies.
Furthermore, there is no sign for a contribution of the \isotope[113m]Cd $\beta$-decay ($Q_\beta=585.7$\,keV, $T_{1/2} = 14.1\,$yr) as considered in Ref.~\cite{Belli2007}.
Since the detectors have been underground at the LNGS since at least 3.5 years (installation of first detectors in Sept.'11, finalized setup since Nov.'13), short-lived cosmogenics affecting the low-energy region decayed away.

The ratio of the integrals over the \cd113 ROI and the total combined spectrum is 99.84\%, indicating again the overwhelming dominance of the \cd113 decay for COBRA.

\section{Analysis}
\subsection{Preparation of templates}
The measured $\beta$-spectra are compared to sets of \cd113 template spectra calculated in different nuclear models in dependence on \gA.
The calculations have been carried out for \mbox{$g_{\rm A} \in [0.8,1.3]$} in 0.01 steps with an energy binning of about 1\,keV.
The upper bound of this range is motivated by the free value of the axial-vector coupling $g_{\rm A}^\text{free} = 1.276(4)$ \cite{UCNA2010}.
In order to compare the data for arbitrary \gA\ values in the given range of the original templates, so-called \textit{splines} are used to interpolate the bin content between different values of \gA\ for each energy bin. 
In contrast to a conventional parameter fit, no optimization process is involved since a spline is uniquely defined as a set of polynomial functions over a range of points $(x_n, y_n)$, referred to as \textit{knots}, and a set of boundary conditions.
Per definition the original templates forming the knots are contained in the spline.
For the spline construction, the \textit{TSpline3} class of the ROOT \cite{Brun1997} software package is used, which utilizes polynomials of grade three.
For the comparison with the data, the finite energy resolution and the electron detection efficiency have to be taken into account.
This is done by folding the templates with the detector-specific energy resolution and the energy dependent detector response function $\varepsilon_\text{det}(E)$.
The latter is determined via a MC simulation assuming an average $xy$-dimension of 10.2\,mm and a height of 10.0\,mm to model the cubic CdZnTe crystals.
It utilizes mono-energetic electrons of starting energies $E_i$, which are homogeneously distributed over the complete volume, and comprises $10^6$ electrons for $E_i \in [4,340]\,$keV in steps of 4\,keV.
The resulting response matrix also takes into account partial energy loss of the electrons and is used to extract $\varepsilon_\text{det}(E)$ in form of a polynomial.
The small deviations observed for the $xy$-dimension of individual crystals are treated as a systematic uncertainty (see section \ref{sec:systematics}).
Nevertheless, these variations are expected to have a rather small effect since the intrinsic detection efficiency $\varepsilon_\text{int}$ for such low energies is very high.
At the $Q$-value of \cd113 it can be quoted as $\varepsilon_\text{int}(Q_\beta) = 97.7\%$ assuming the average crystal size.
For lower energies the efficiency is continuously increasing.

Ref. \cite{Belli2007} used a simplified approach to correct for the efficiency of their CdWO$_4$ scintillator setup and introduced an energy independent scaling factor of $\varepsilon=99.97\%$.
This might affect the spectrum shape at low energies.

Finally, each convolved template spectrum is normalized by the integral over the accessible energy range depending on the threshold of each individual detector.

\subsection{Spectrum shape comparison}
\label{sec:shape_comparison}
As the individual detector thresholds had to be adjusted slightly over the run time of the \cd113\ data-taking, it is necessary to normalize each energy bin of the experimental spectra with its corresponding exposure.
The bin width is set to 4\,keV, which is a compromise between a large number of bins $N$ \mbox{-- beneficial} for the anticipated $\chi^2$ test to compare the spectrum shapes -- and the assigned bin uncertainties arising from the number of entries per bin.
The fixed binning requires to remove the lowest bin of each spectra, because $E_\text{th}$ is not necessarily a multiple of 4\,keV.
Additionally, as a conservative approach to address that some noise contribution might still be leaking into the \cd113 spectra for certain periods where the thresholds had to be increased along with a potential signal loss due to the DCC efficiency, the analysis threshold is set to $\tilde{E}_\text{th} = E_\text{th} + 8$\,keV.
This also increases the average threshold $\overline{E}_\text{th}$ to 91.9\,keV.
Following, the experimental spectra are normalized to unity as well.

In contrast to this conservative approach, Ref. \cite{Belli2007} used a finer binning, while their resolution is at least two times worse.
The noise influence was corrected using a pulse-shape discrimination technique along with the deduced signal efficiency, which was estimated from calibration data, but no further threshold was introduced in this study.

\begin{figure}[ht!]
\centering
\includegraphics[width=0.5\textwidth]{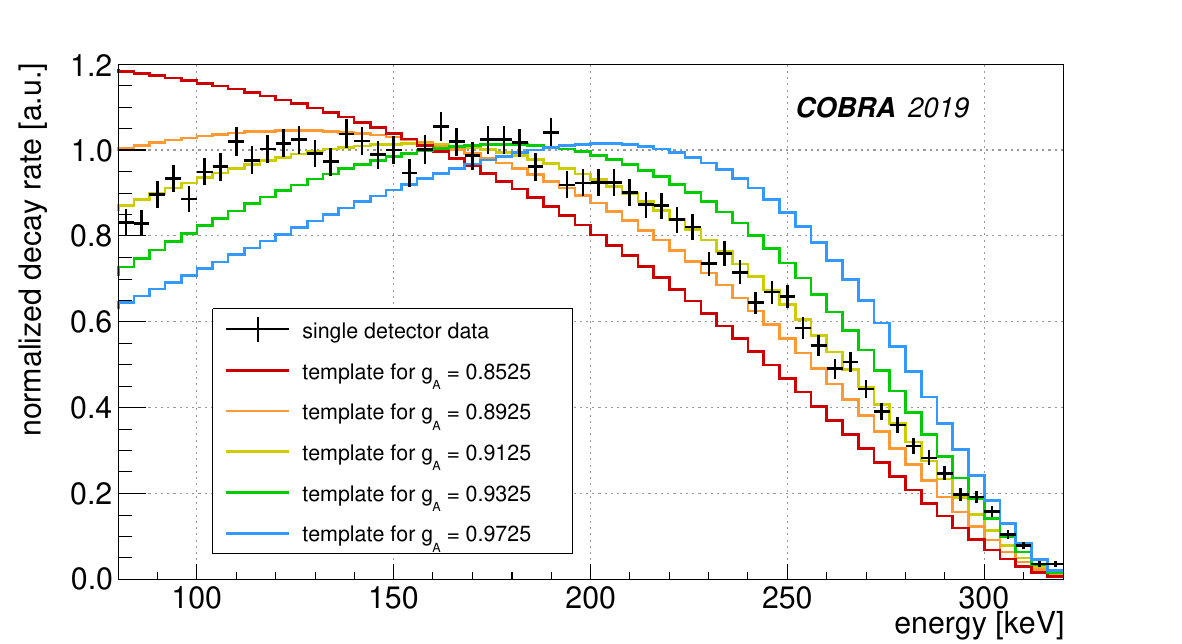}
\caption{Comparison of five interpolated, normalized template spectra based on the ISM calculations for the \cd113 $\beta$-electron distribution and one COBRA single detector spectrum.}
\label{fig:shape_comparison}
\end{figure}

\begin{figure}[ht!]
\centering
\includegraphics[width=0.5\textwidth]{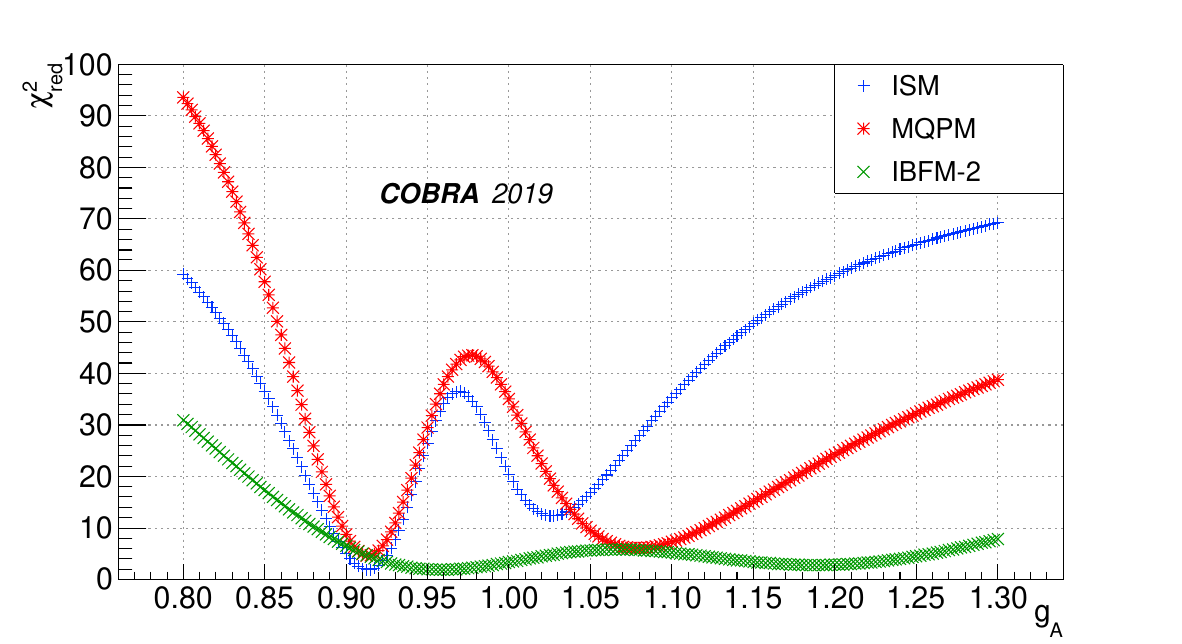}
\caption{$\chi^2_\text{red}(g_{\rm A})$ curve for the spectrum shape comparison of one COBRA single detector spectrum and the interpolated templates provided by the ISM, MQPM and IBFM-2. The shape of the $\chi^2_\text{red}(g_{\rm A})$ curves presented here is representative for the complete detector ensemble and the combined spectrum.}
\label{fig:chis2}
\end{figure}

Using the experimental values $m_i$ of the energy bins $i$ to $N$ with Poisson uncertainties $\sigma_i$ and the corresponding prediction $t_i$ based on the template calculated for a certain \gA , the quantity $\chi^2$ is derived as
\begin{eqnarray}
\chi^2 = \sum \limits_{i=1}^N \left( \frac{m_i - t_i}{\sigma_i} \right)^2 .
\end{eqnarray}
A comparison between one of the single detector measurements and a subset of interpolated \cd113 ISM templates is illustrated in Fig.~\ref{fig:shape_comparison}.
For the same detector the reduced \mbox{$\chi^2_\text{red}(g_{\rm A}) = \chi^2(g_{\rm A})/(N-1)$} in the given \gA\ range is shown in Fig.~\ref{fig:chis2} for the three nuclear model calculations.
This procedure is repeated for all 45 independent detector spectra to extract the best match \gA\ value from the minimum of the $\chi^2_\text{red}(g_{\rm A})$ curve with a parabola fit for each of the models.
The uncertainty on every best match \gA\ is derived from the minimum $\chi^2 + 1$ as $1\sigma$ deviation.
Additionally, the analysis is performed for the combination of the individual spectra using average values to convolve the templates.
A compilation of all the experimental spectra with the final threshold $\tilde{E}_\text{th}$ can be found in the appendix (see Fig.~\ref{fig:appendix_single_spectra01}\,--\,\ref{fig:appendix_single_spectra05}).

\subsection{Results}
\begin{figure}[ht!]
\centering
\includegraphics[width=0.5\textwidth]{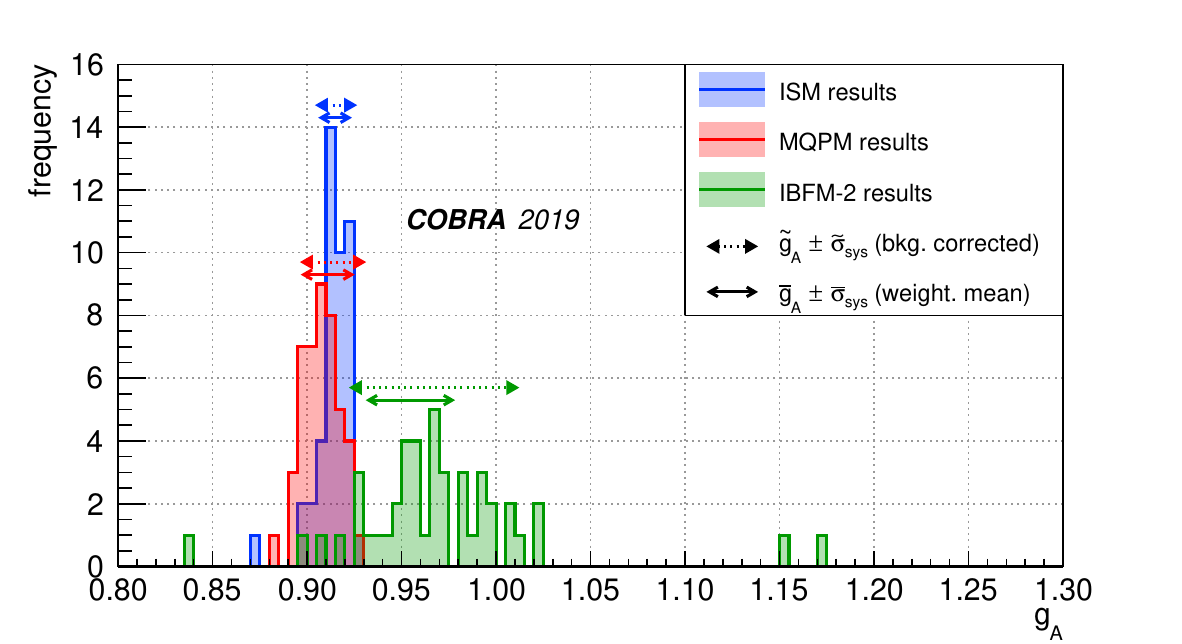}
\caption{Distribution of the 45 best match \gA\ values for the ISM, MQPM and IBFM-2. Additionally, the weighted mean $\overline{g}_{\rm A} \pm \overline{\sigma}_\text{sys}$ as well as the result of the spectrum shape comparison for the combined spectrum $\tilde{g}_{\rm A} \pm \tilde{\sigma}_\text{sys}$ including the background correction are highlighted.}
\label{fig:gA_distribution}
\end{figure}

The resulting distributions of the best match \gA\ values for the 45 independent measurements and the three nuclear models considered are shown in Fig.~\ref{fig:gA_distribution}.
While the ISM and MQPM results are tightly distributed around a common mean value, the IBFM-2 distribution is much wider.
This is due to the fact that the latter model is less sensitive to \gA\ as can also be seen in the $\chi^2_\text{red}(g_{\rm A})$ curve in Fig. \ref{fig:chis2}.

From those single detector results a weighted mean using the $\chi^2+1$ deviation can be constructed to extract an average $\overline{g}_{\rm A}$ for each model.
The statistical uncertainty $\overline{\sigma}_\text{stat}$ on $\overline{g}_{\rm A}$ turns out to be negligibly small considering the systematics $\overline{\sigma}_\text{sys}$ as done in section~\ref{sec:systematics}.
They are on the order of $\overline{\sigma}_\text{stat} \sim 2\cdot 10^{-4}$ for ISM and MQPM and a factor of four higher for IBFM-2, respectively.
The extracted weighted means including the dominant systematic uncertainties yield the following results
\begin{eqnarray}
\overline{g}_{\rm A}(\text{ISM})    &=& 0.915 \pm 0.007, \label{eqn:ISM_result}\\
\overline{g}_{\rm A}(\text{MQPM})   &=& 0.911 \pm 0.013,\\
\overline{g}_{\rm A}(\text{IBFM-2}) &=& 0.955 \pm 0.022. \label{eqn:IBFM_result}
\end{eqnarray}

These values are in perfect agreement with the results obtained for the combined spectrum, where the MC prediction as presented in section \ref{sec:background} is used to correct for the underlying background (see Fig.~\ref{fig:gA_distribution}).
For the single detector analysis the background model is not used explicitly, but it enters as systematic uncertainty as discussed in the next section.

\subsection{Systematic uncertainties}
\label{sec:systematics}
The systematic uncertainties are determined after fixing all input parameters of the spectrum shape analysis.
They are evaluated separately by modifying one considered parameter within conservative limits while the other parameters are fixed to their default values in the analysis.
The modulus of the difference between the altered and the default \gA\ result is then taken as a measure for the systematic uncertainty. 
The total systematic uncertainty for each model is obtained as the square root of the sum of squared uncertainties. A summary of the systematics is given in Tab.~\ref{tab:systematics}.

\begin{table}[ht!]
\caption{Summary of systematic uncertainties.}
\label{tab:systematics}
\begin{tabular}{ccccc}
\toprule
          &  & Uncertainty [\%] &  \\
Parameter & ISM & MQPM & IBFM-2\\ 
\midrule
efficiency $\varepsilon(E)$   & 0.010 & 0.011 & 0.033\\
resolution FWHM$(E)$          & 0.060 & 0.032 & 0.226\\
energy calibration		      & 0.751 & 0.796 & 2.130\\
threshold				      & 0.131 & 1.120 & 0.714\\
$z$-cut selection		      & 0.120 & 0.192 & 0.457\\
template interpolation        & 0.002 & 0.002 & 0.001\\
$\chi^2_\text{red}$ fit range & 0.102 & 0.058 & 0.034\\
background modeling			  & 0.042 & 0.016 & 0.068\\
\midrule
total 						  & 0.798 & 1.389 & 2.306\\
\bottomrule
\end{tabular}
\end{table}

The effect of the efficiency scaling is studied by changing the crystal size in the MC simulation to the minimum and maximum physical $xy$-dimensions of the selected detectors.
The systematic differences are then added in quadrature.
To evaluate the influence of the resolution smearing, the FWHM$(E)$ is fixed to the worst and best resolution curve.
Following, the spectrum shape analysis is repeated with fixed FWHM$(E)$ and the systematic differences are again added in quadrature.
The influence of a misaligned energy calibration is studied by shifting the experimental data according to the uncertainty of the calibration and add the systematic differences in quadrature.
An average peak shift of $\pm$1.3\,keV was found for the 238.6\,keV line of the combined \isotope[228]Th calibration, allowing to neglect the uncertainty on the accepted \cd113 $Q$-value quoted in section \ref{sec:previous_Cd113}.
It turns out that the calibration uncertainty is one of the dominating contributions.
Increasing the analysis threshold for all detectors to at least 120\,keV, which is more than two FWHM on average, is taken as a measure for systematic effects due to the threshold optimization and individual values.
The effect is different for the nuclear models and ensues from the change of the predicted shape at low energies for MQPM compared to the other two models and the weaker \gA\ dependence of the IBFM-2 calculations.
The systematic uncertainty of the $z$-cut selection is evaluated by slicing each detector in two disjunct depth ranges to perform the analysis for both slices independently.
The systematic differences are again added in quadrature.
The accuracy of the spline interpolation is evaluated by removing the template that is the closest to the best match \gA\ from the given ensemble and determining the difference between the reduced and the full spline result.
This is a conservative approach since the removed template is part of the final spline. 
The influence of the fit range to extract the minimum of $\chi^2_\text{red}(g_{\rm A})$ is taken into account as another systematic uncertainty.
Finally, the effect of neglecting the average background model in the single detector analysis is inferred from the analysis results of the combined spectrum with and without background subtraction.
As expected from the superb $S/B$ ratio, the effect is only marginal, which justifies the procedure.

Additional systematics as considered in previous studies (e.g. \cite{Cobra_Cd113_2009, Belli2007}), like the exact amount of \cd113 in the crystals, potential dead layer effects or a varying composition of CdZnTe due to the complex crystal growth, do not have an influence on the spectrum shape analysis.
Those effects are only important for extracting the total decay rate, which is needed to determine the decay's half-life.

In total, the systematic uncertainties add up to values on the percent level and agree well with the observed spread of the $g_{\rm A}$ distributions as seen in Fig.~\ref{fig:gA_distribution}.
Furthermore, the single detector results are consistent with the analysis results of the combined spectrum, which yields about 30\% higher uncertainties.

\subsection{Discussion}

\begin{figure*}[ht!]
\centering
\includegraphics[width=0.33\textwidth]{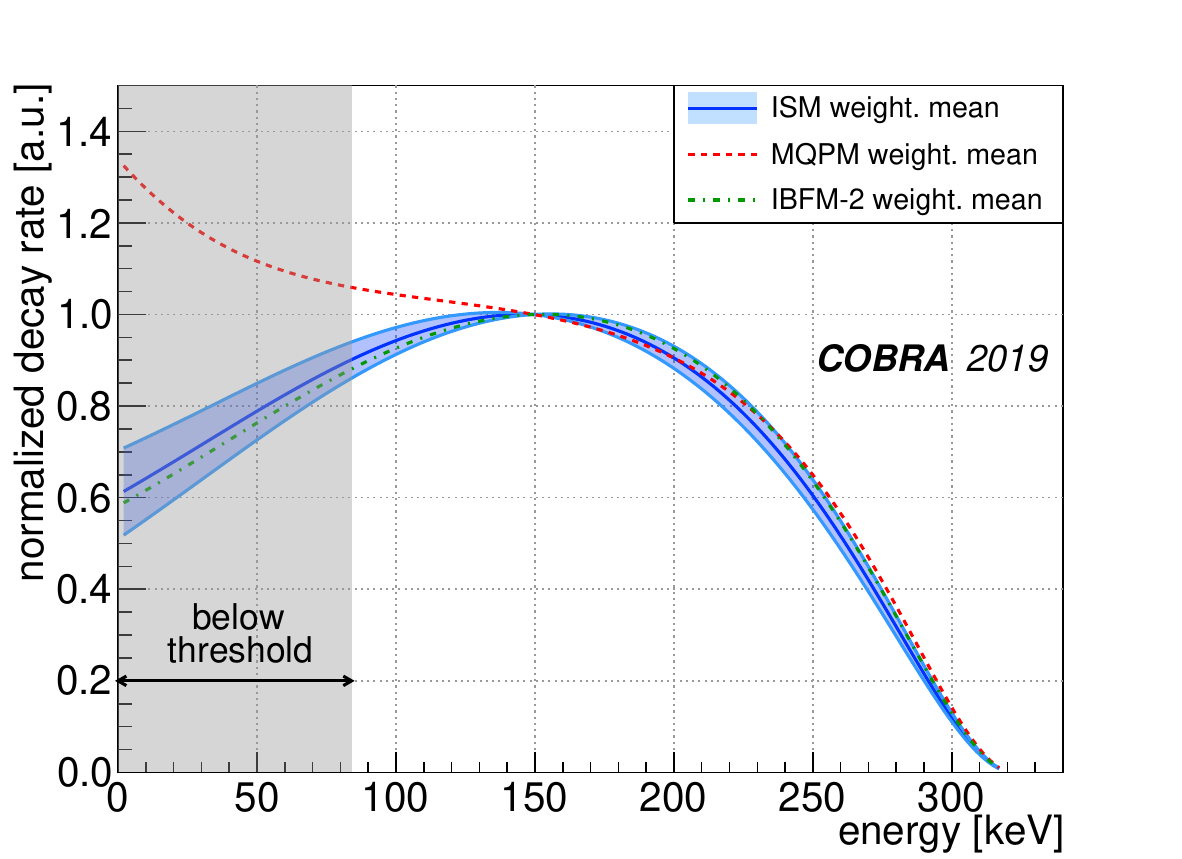} \includegraphics[width=0.33\textwidth]{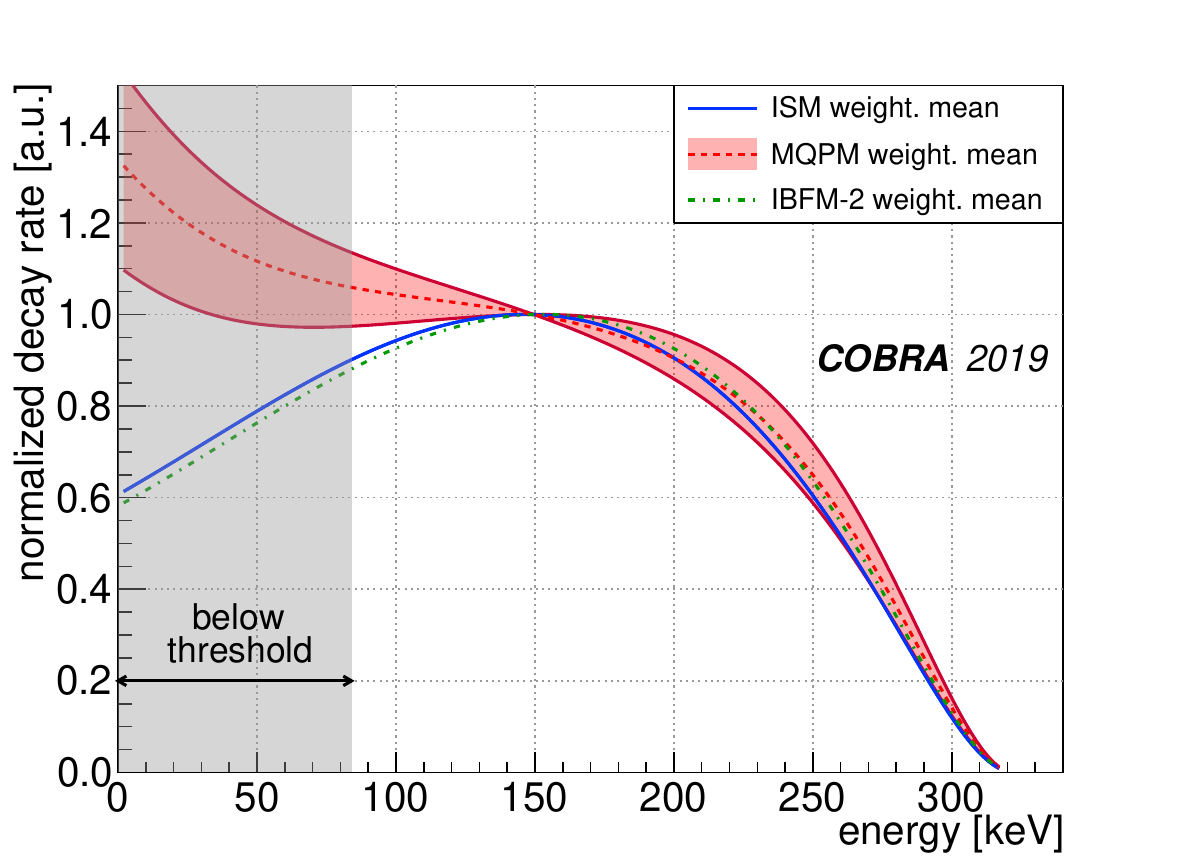}
\includegraphics[width=0.33\textwidth]{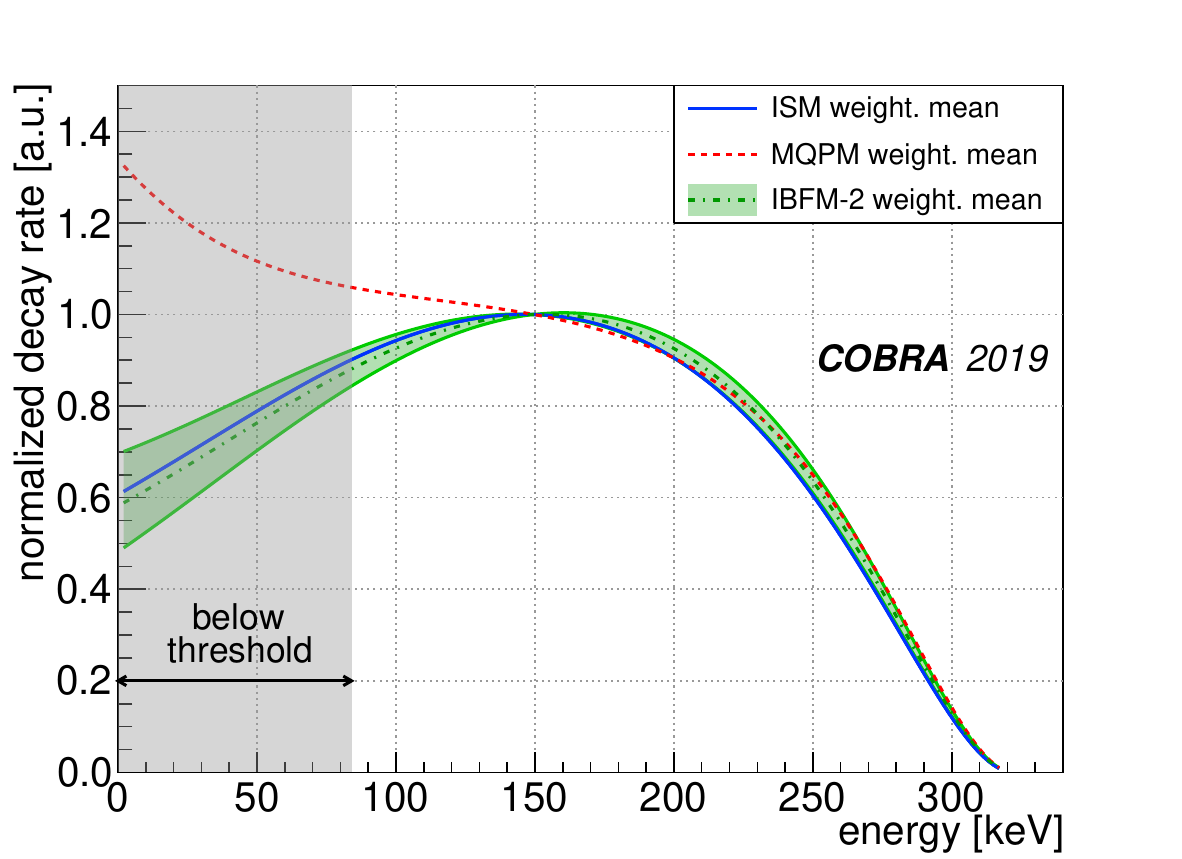}
\caption{Allowed spectrum range for the \cd113 $\beta$-decay according to the ISM (left), MQPM (middle) and IBFM-2 (right) templates interpolated for the determined $\overline{g}_{\rm A} \pm \overline{\sigma}_\text{sys}$ including the experimental uncertainties. For comparison, the template corresponding to the weighted mean is shown for the respective other models.}
\label{fig:allowed_spectrum_range}
\end{figure*}

\begin{figure*}[ht!]
\centering
\includegraphics[width=0.33\textwidth]{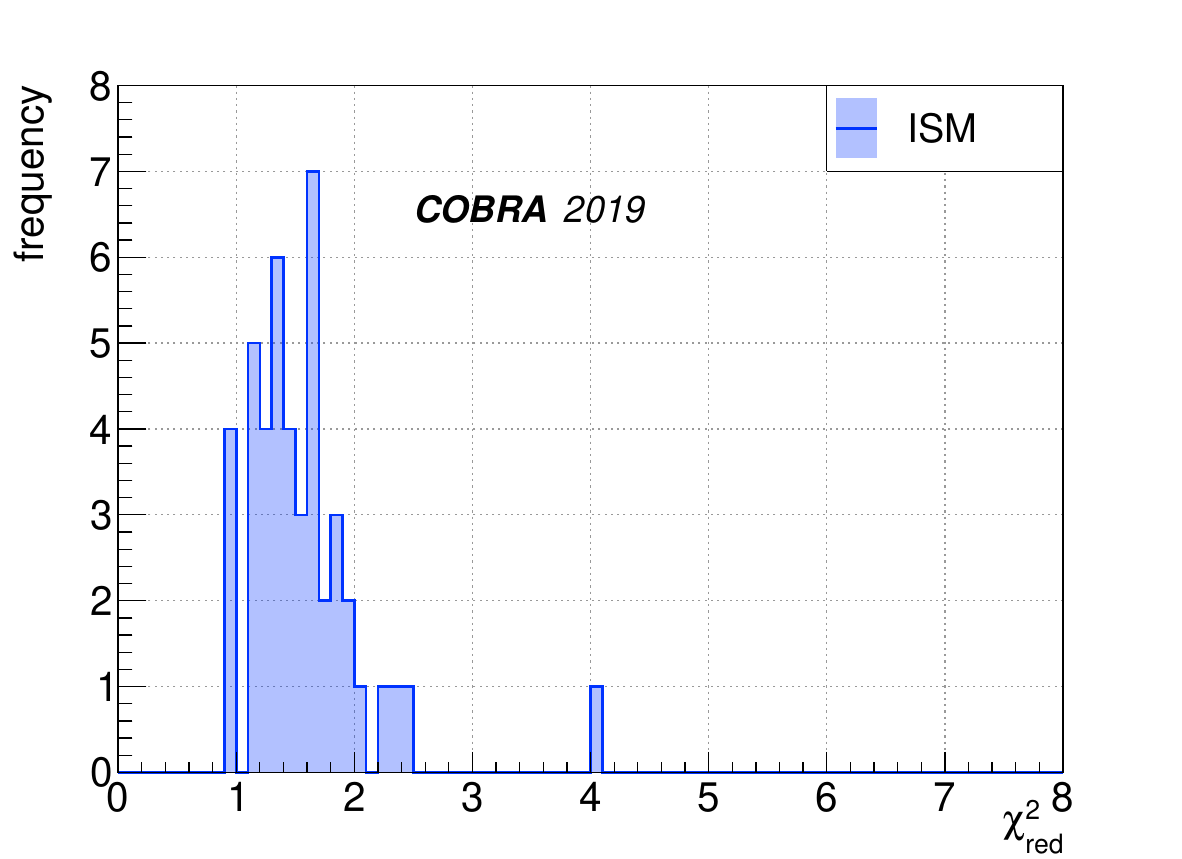}
\includegraphics[width=0.33\textwidth]{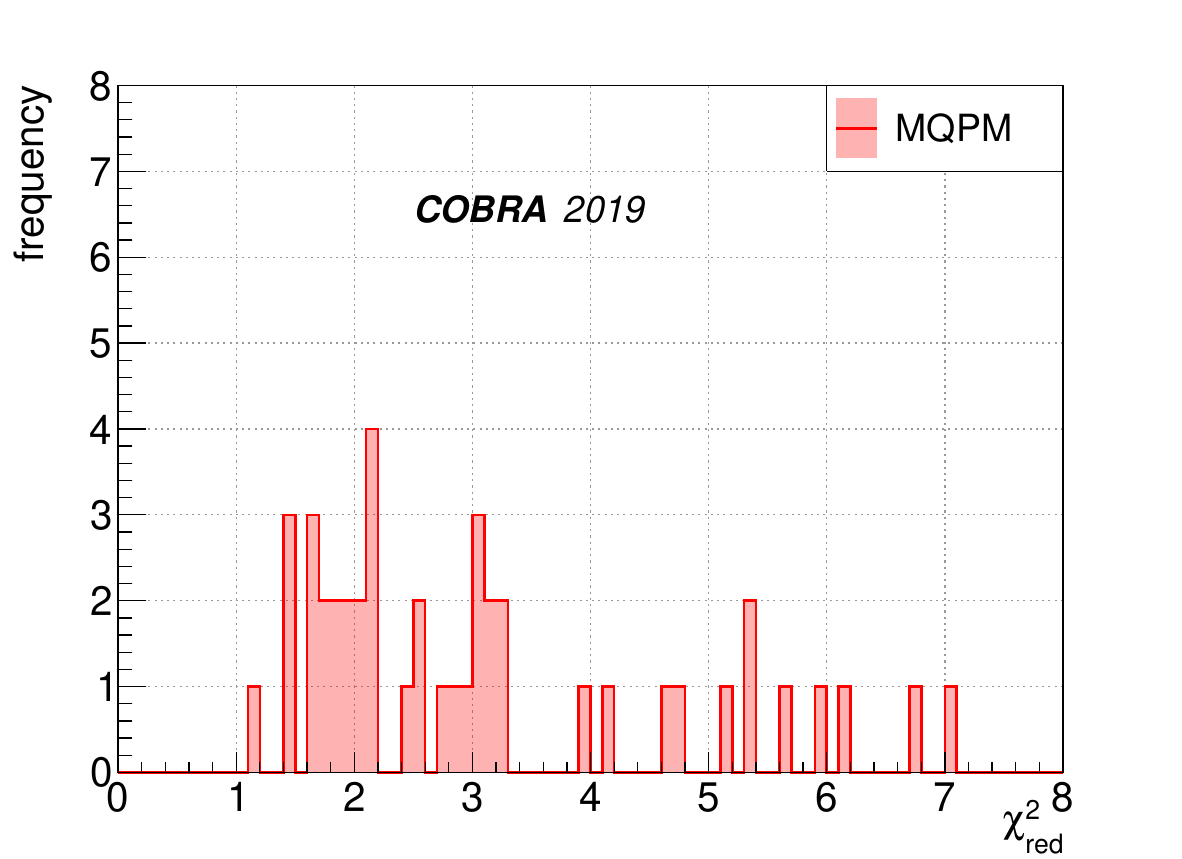}
\includegraphics[width=0.33\textwidth]{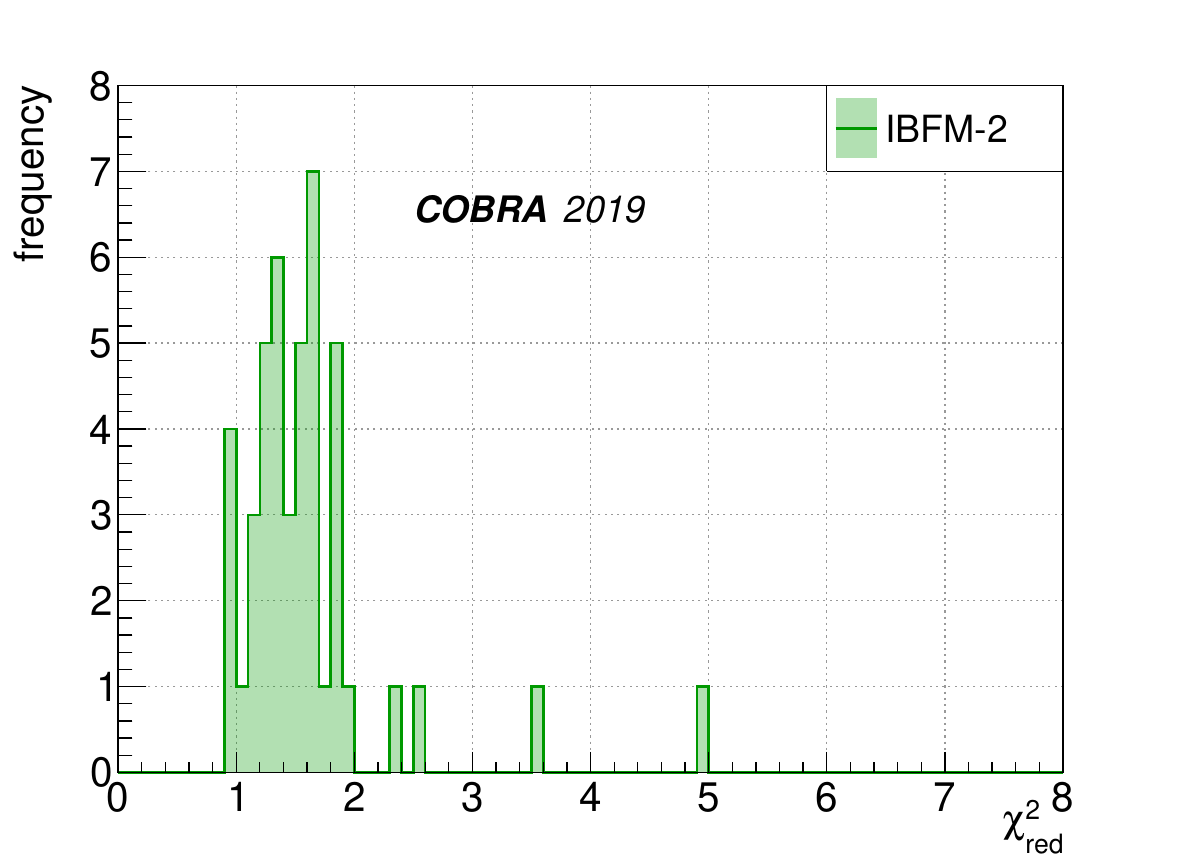}
\caption{Distribution of the minimum $\chi^2_\text{red}$ values of the 45 best match \gA\ values for the single detector spectrum shape comparison.}
\label{fig:chi2_distributions}
\end{figure*}

The average \gA\ values in combination with the determined experimental uncertainties can be used to illustrate the allowed spectrum range for the \cd113 $\beta$-decay using the matching interpolated templates without incorporating detector effects such as finite energy resolution and efficiency (see Fig.~\ref{fig:allowed_spectrum_range}).

While the spectral shape is very similar for the ISM and \mbox{IBFM-2}, the trend at low energies is contrary for the MQPM prediction.
Nonetheless, the spectral shapes seem to be in good agreement for energies above 100\,keV for all three models.
Even though the IBFM-2 is associated with the highest experimental uncertainty, a comparable allowed spectrum range is achieved due to the fact that the model is less sensitive to \gA.
Fig.~\ref{fig:chi2_distributions} shows the minimum $\chi^2_\text{red}$ distributions corresponding to the 45 best match \gA\ values.
Again, the results based on the ISM and IBFM-2 calculations are very similar. 
The average values $\overline{\chi}^2_\text{red}(\text{ISM}) = 1.57 \pm 0.08$, $\overline{\chi}^2_\text{red}(\text{MQPM}) = 3.27 \pm 0.28$ and $\overline{\chi}^2_\text{red}(\text{IBFM-2}) = 1.62 \pm 0.10$ considering the uncertainty of the mean of the distributions, indicate that there is less agreement between the MQPM prediction and the experimentally observed spectrum shape than for ISM and IBFM-2.
This is why there is a slight preference for the ISM prediction due to the tightly distributed single detector results, the assigned systematic uncertainty and the pleasing minimum $\chi^2_\text{red}$ distribution.

\section{Conclusion}
The spectrum shape of the fourfold forbidden non-unique \mbox{$\beta$-decay} of \cd113 has been investigated with 45 CdZnTe detectors and an average analysis threshold of 91.9\,keV.
The data set corresponds to an isotopic exposure of 2.89\,kg\,d.
Each individual \cd113 $\beta$-spectrum was evaluated in the context of three nuclear models to extract average values of the effective axial-vector coupling \gA.
The data support the idea that \gA\ is quenched in the low-momentum-exchange $\beta$-decay of \cd113 independently of the underlying nuclear model.
Nevertheless, the low-energy region needs to be explored further to distinguish the contrary behavior of the spectrum shape predicted by the different models.

\section*{Acknowledgments}
We thank the LNGS for the continuous support of the \mbox{COBRA} experiment, and the AMANDA collaboration, especially R. Wischnewski, for providing us with their FADCs.
\mbox{COBRA} is supported by the German Research Foundation DFG (ZU123/3 and GO1133/1) and by the Ministry of Education, Youth and Sports of the Czech Republic with contract no. CZ.02.1.01/0.0/0.0/16\_013/0001733.
Additionally, support from the Academy of Finland under the project no. 318043 and from the Jenny and Antti Wihuri Foundation is acknowledged.

\bibliographystyle{unsrt}
\bibliography{Cobra_Cd113-gA-study_2019_PLB}



\appendix
\section{Single detector spectra}

\begin{figure*}
\centering
\includegraphics[width=0.9\textwidth]{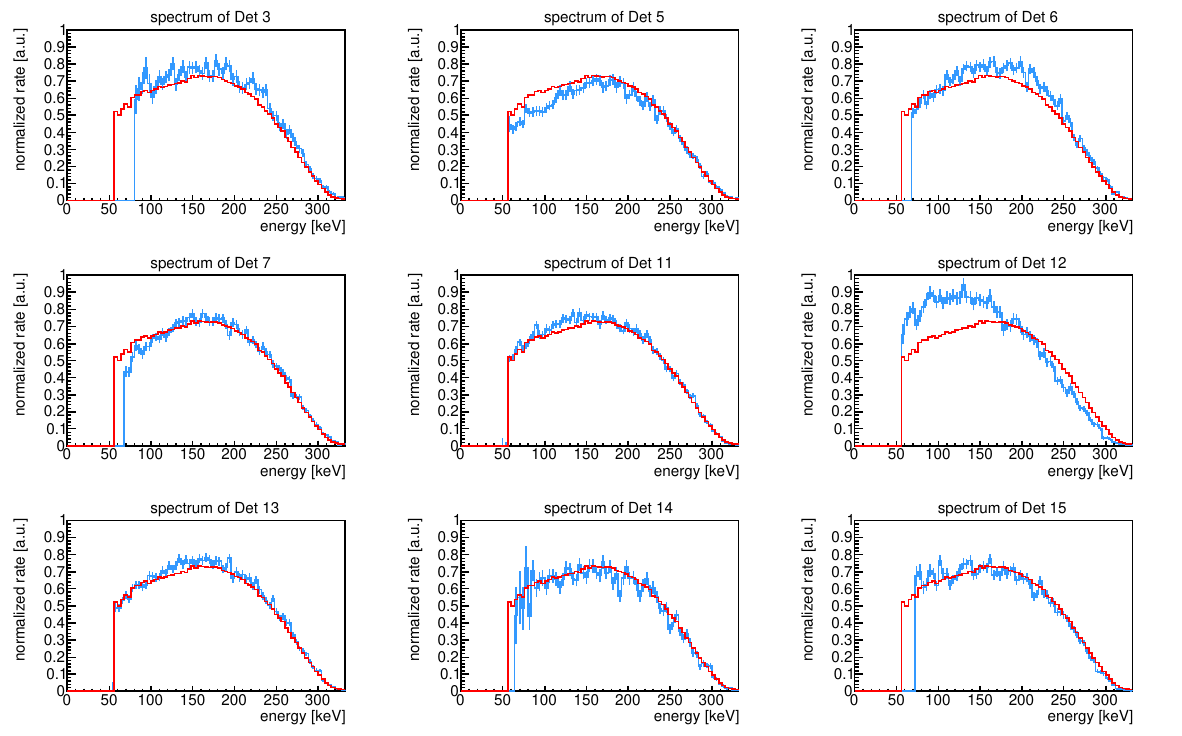}
\caption{Compilation of the experimental \cd113 spectra measured with the COBRA demonstrator (Det\,3 -- Det\,15). Each spectrum is normalized by the integral from the respective threshold to the $Q$-value of \cd113. The combination of all detectors is shown for comparison (red solid histogram).}
\label{fig:appendix_single_spectra01}
\end{figure*}

\begin{figure*}
\centering
\includegraphics[width=0.9\textwidth]{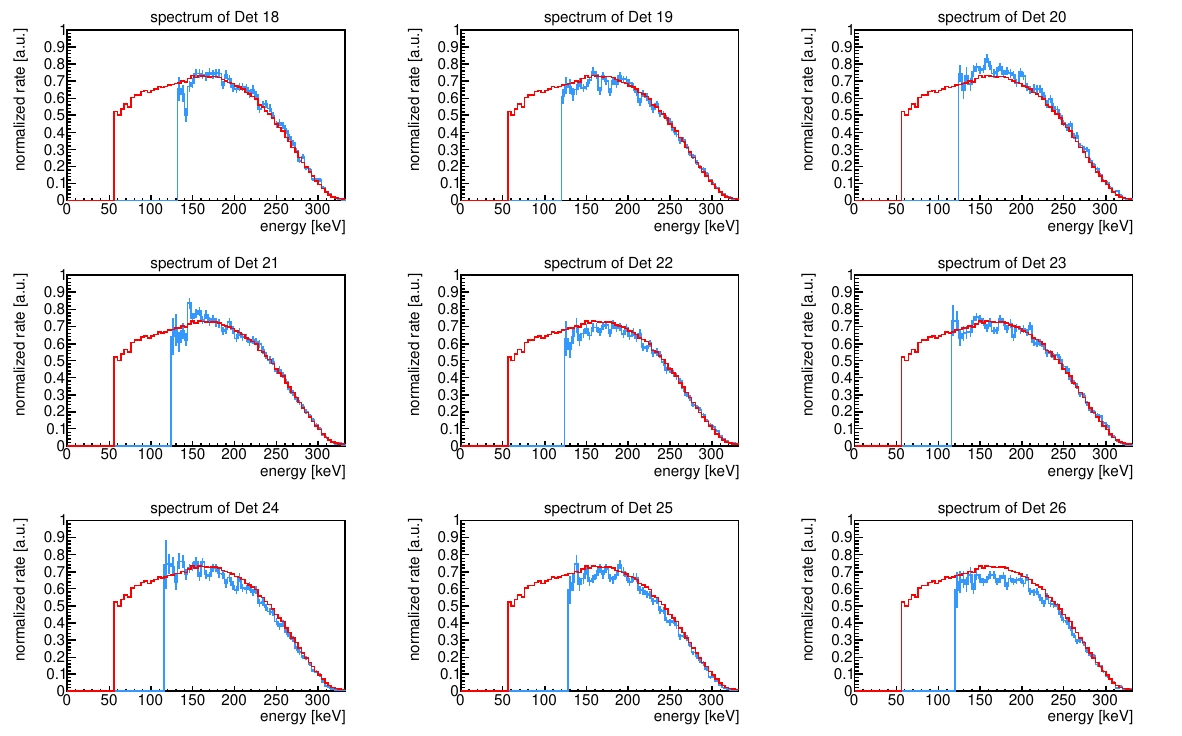}
\caption{Compilation of the experimental \cd113 spectra measured with the COBRA demonstrator (Det\,18 -- Det\,26). Each spectrum is normalized by the integral from the respective threshold to the $Q$-value of \cd113. The combination of all detectors is shown for comparison (red solid histogram).}
\label{fig:appendix_single_spectra02}
\end{figure*}

\begin{figure*}
\centering
\includegraphics[width=0.9\textwidth]{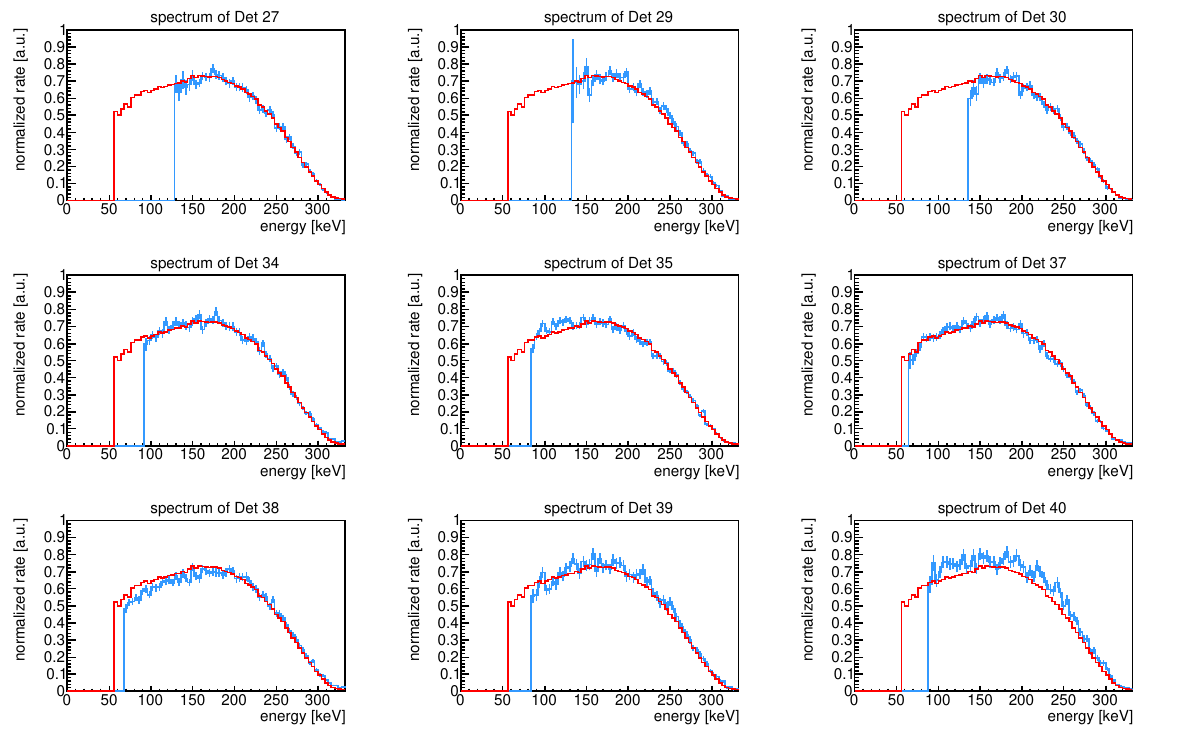}
\caption{Compilation of the experimental \cd113 spectra measured with the COBRA demonstrator (Det\,27 -- Det\,40). Each spectrum is normalized by the integral from the respective threshold to the $Q$-value of \cd113. The combination of all detectors is shown for comparison (red solid histogram).}
\label{fig:appendix_single_spectra03}
\end{figure*}

\begin{figure*}
\centering
\includegraphics[width=0.9\textwidth]{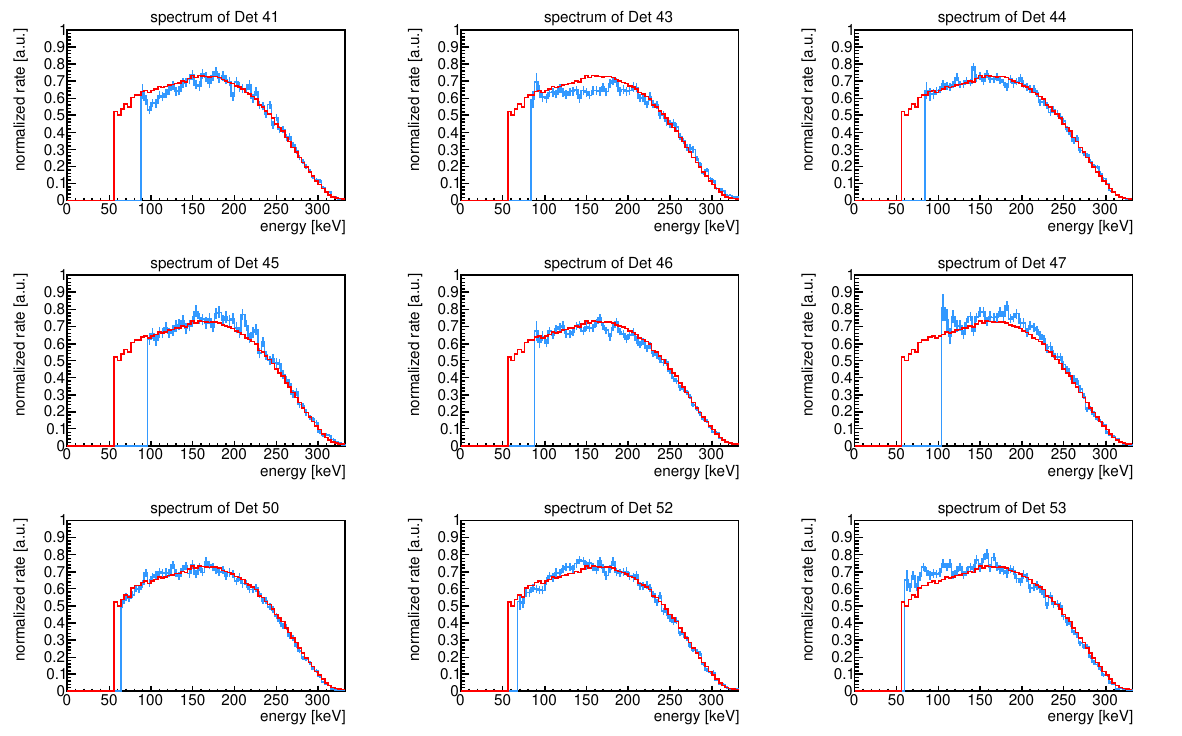}
\caption{Compilation of the experimental \cd113 spectra measured with the COBRA demonstrator (Det\,41 -- Det\,53). Each spectrum is normalized by the integral from the respective threshold to the $Q$-value of \cd113. The combination of all detectors is shown for comparison (red solid histogram).}
\label{fig:appendix_single_spectra04}
\end{figure*}

\begin{figure*}
\centering
\includegraphics[width=0.9\textwidth]{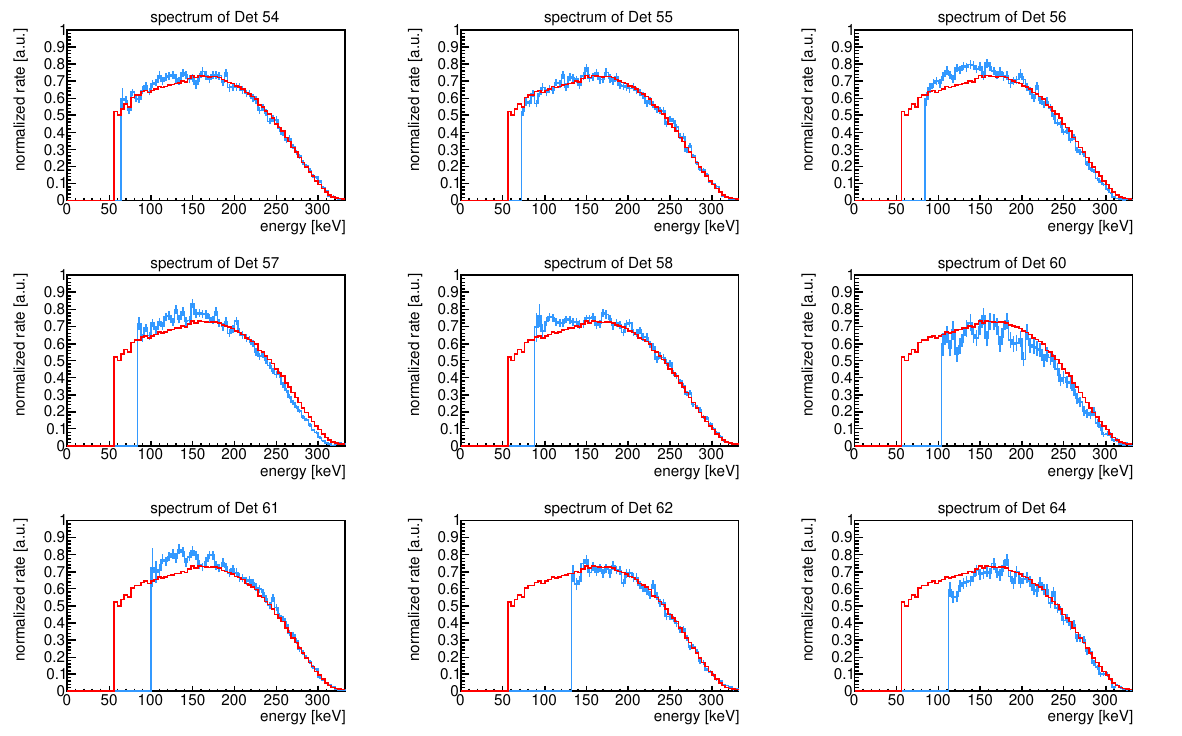}
\caption{Compilation of the experimental \cd113 spectra measured with the COBRA demonstrator (Det\,54 -- Det\,64). Each spectrum is normalized by the integral from the respective threshold to the $Q$-value of \cd113. The combination of all detectors is shown for comparison (red solid histogram).}
\label{fig:appendix_single_spectra05}
\end{figure*}





\end{document}